# Static behaviour of induced seismicity

A. Mignan

Institute of Geophysics, Swiss Federal Institute of Technology Zurich, Switzerland

arnaud.mignan@sed.ethz.ch

**Abstract:** The standard paradigm to describe seismicity induced by fluid injection is to apply nonlinear diffusion dynamics in a poroelastic medium. I show that the spatiotemporal behaviour and rate evolution of induced seismicity can, instead, be expressed by geometric operations on a static stress field produced by volume change at depth. I obtain laws similar in form to the ones derived from poroelasticity while requiring a lower description length. Although fluid flow is known to occur in the ground, it is not pertinent to the behaviour of induced seismicity. The proposed model is equivalent to the static stress model for tectonic foreshocks generated by the Non-Critical Precursory Accelerating Seismicity Theory. This study hence verifies the explanatory power of this theory outside of its original scope.

## 1. Introduction

Induced seismicity is a growing concern for the energy-sector industry relying on fluid injection in the deep parts of the Earth's crust [Ellsworth, 2013; Mignan et al., 2015]. At the same time, fluid injection sites provide natural laboratories to study the impact of increased fluid pressure on earthquake generation [Majer et al., 2007]. Induced seismicity is characterised by two empirical laws, namely (*i*) a linear relationship between the fluid mass $m(t)$ injected up to time $t$ and the cumulative number of induced earthquakes $N(t)$ and (*ii*) a parabolic induced seismicity envelope radius $r \propto \sqrt[n]{m(t)}$ with $n$ a positive integer [Shapiro and Dinske, 2009]. These two descriptive laws can be derived from the differential equations of poroelasticity [Biot, 1941] under various assumptions. The full description of the process requires complex numeric modelling coupling fluid flow, heat transport and geomechanics [Rutqvist, 2011]. These models, numerically cumbersome, can become intractable because of the sheer number of parameters [Miller, 2015]. Attempts to additionally correct for the known discrepancies between Biot's theory and rock experiments have led to a large variety of model assumptions [Berryman and Wang, 2001], indicating that poroelasticity results are ambiguous.

I will demonstrate that a simple static stress model can explain the two empirical laws of induced seismicity without requiring any concept of poroelasticity. The proposed theoretical framework hence avoids the aforementioned shortcomings by suggesting an origin of induced seismicity that does not involve fluid flow in a porous medium (although fluid flow indeed occurs). Historically, a similar static stress model was proposed for the tectonic regime under the Non-Critical Precursory Accelerating Seismicity Theory (N-C PAST) [Mignan et al., 2007; Mignan, 2012]. Its application to induced seismicity data will allow a more fundamental investigation of the relationship between static stress and earthquake generation. To test the model, I will use data from the 2006 Basel Enhanced Geothermal System (EGS) stimulation experiment including the flow rate of injected fluids [Häring et al., 2008] and the relocated catalogue of induced seismicity [Kraft and Deichmann, 2014].

## 2. The Non-Critical Precursory Accelerating Seismicity Theory (N-C PAST)

The N-C PAST has been proposed to explain the precursory seismicity patterns observed before large earthquakes from geometric operations in the spatiotemporal stress field generated by constant tectonic stress accumulation [Mignan et al., 2007; Mignan, 2012]. In particular, it provides a mathematical expression of temporal power-laws without requiring local interactions between the elements of the system [Sammis and Sornette, 2002; Mignan, 2011]. Therefore earthquakes are considered passive (static) tracers of the stress accumulation process, in contrast with active earthquake cascading in a critical process (hence the term "non-critical"). The concept of self-organized criticality [Bak and Tang, 1989] is seldom used to explain induced seismicity [Grasso and Sornette, 1998]. Since there is no equivalent of a mainshock in induced seismicity, the criticality versus non-criticality debate has limited meaning in that case. However, the underlying process of static stress changes considered in the N-C PAST can be tested against the observed spatiotemporal behaviour of induced seismicity.

The N-C PAST postulates that earthquake activity can be categorized in three regimes – background, quiescence and activation – depending on the spatiotemporal stress field $\sigma(r,t)$

$$\sigma(r,t) = \begin{cases} \sigma_0^* & , t < t_0 \\ \frac{h^n}{(r^2+h^2)^{\frac{n}{2}}}\left(\sigma_0 + \dot{\tau}(t-t_0)\right) + \sigma_0^* & , t_0 \leq t < t_f \end{cases} \qquad (1)$$

defined from the boundary conditions $\sigma(r \rightarrow +\infty, t) = \sigma_0^*$ and $\sigma(r = 0, t) = \sigma_0 + \dot{\tau}t + \sigma_0^*$, with $h$ the depth of the fault segment base, $r$ the distance along the stress field gradient from the fault's surface projection, $\sigma_0 < 0$ the stress drop associated to a hypothetical silent slip occurring at $t_0$ at the base of the fault, $\dot{\tau}$ the tectonic stress rate on the fault, $\sigma_0^*$ the crustal background stress, $n = 3$ the spatial diffusion exponent for static stress and $t_f$ the mainshock occurrence time [Mignan et al., 2007] (Fig. 1a). Background, quiescence and activation regimes are defined by event densities $\delta_{b0}$, $\delta_{bm}$, and $\delta_{bp}$ for $|\sigma| \leq \sigma_0^* \pm \Delta\sigma^*$, $\sigma < \sigma_0^* - \Delta\sigma^*$ and $\sigma > \sigma_0^* + \Delta\sigma^*$, respectively, with the boundary layer $\pm\Delta\sigma^*$ the background stress amplitude range. By definition, $\delta_{bm} < \delta_{b0}$

< $\delta_{bp}$ with each seismicity regime assumed isotropic and homogeneous in space (i.e. role of fault network neglected). Correlation between earthquake productivity and static stress changes is well established [King, 2007]. The distinction of three unique seismicity regimes with constant event density, the main assumption of the N-C PAST, is discussed later on.

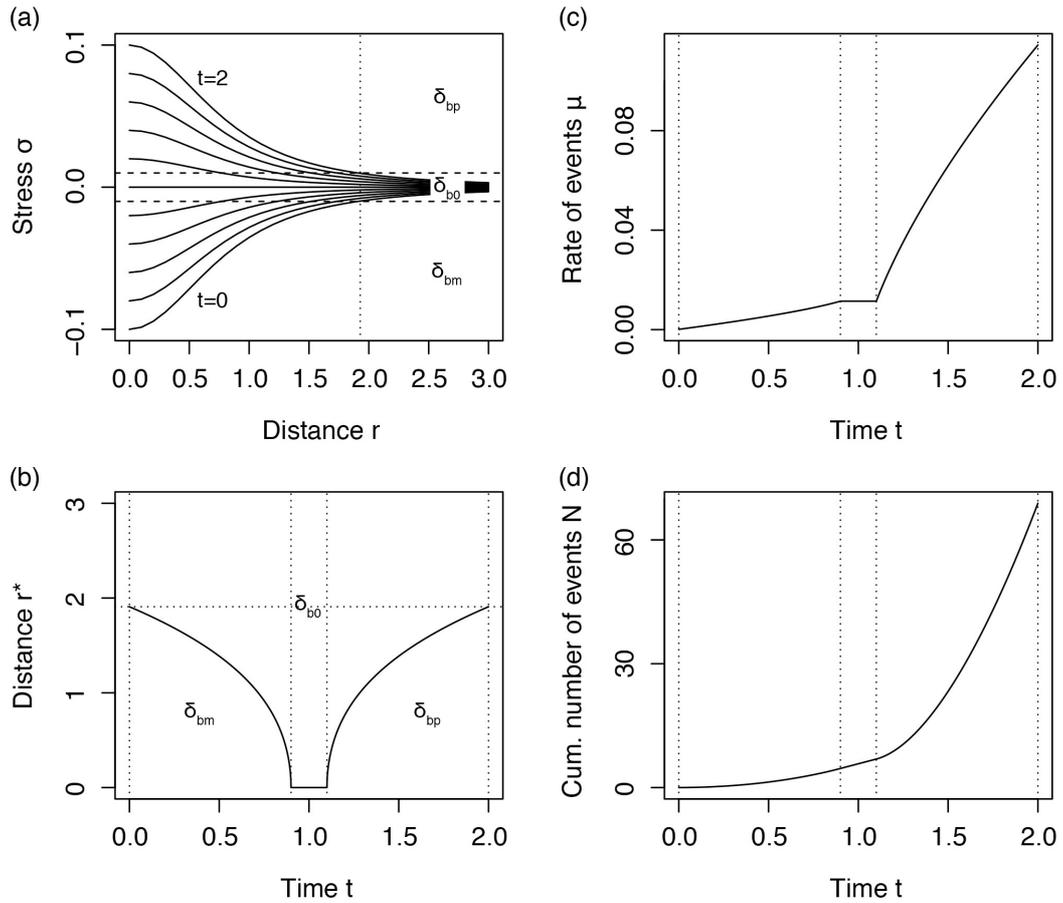

**Figure 1:** Seismicity spatiotemporal behaviour described by the N-C PAST static stress model (tectonic case [Mignan, 2012]): (a) Spatiotemporal evolution of the stress field $\sigma(r,t)$ generated by constant stress accumulation $\dot{\tau}$ on a fault located at $r = 0$ (Eq. 1). Background, quiescence and activation seismicity regimes are described by densities of events $\delta_{b0}$, $\delta_{bm}$, and $\delta_{bp}$ for $|\sigma| \leq \sigma_0^* \pm \Delta\sigma^*$, $\sigma < \sigma_0^* - \Delta\sigma^*$ and $\sigma > \sigma_0^* + \Delta\sigma^*$, respectively; (b) Temporal evolution of quiescence and activation envelopes $r^*(t)$

with σ($r^*$) = $σ_0^* ± Δσ^*$ (Eq. 2); (c) Rate of events μ(*t*) in a disc of constant radius max(*r**) (Eq. 3); (d) Cumulative number of events *N*(t) (Eq. 4) of power-law form (Eq. 5). With $t_0$ = 0, $t_{mid}$ = 1, $t_f$ = 2, *h* = 1, $\dot{τ}$ = 0.1, $σ_0^*$ = 0, $Δσ^*$ = $10^{-2}$, $δ_{bm}$ = 0.001, $δ_{b0}$ = 0.1, $δ_{bp}$ = 1, *n* = 3, *k* = π, *d* = 2, Δ*t* = 0.01.

In the tectonic case, static stress changes are underloading due to hypothetical precursory silent slip on the fault at $t_0$ followed by overloading due to hypothetical asperities delaying rupture on the fault after $t_p^*$ [Mignan, 2012]. The three seismicity regimes are then defined as solid spatiotemporal objects with envelopes

$$\begin{cases} r_Q^*(t_0 \leq t < t_m^*) = h\left[\left(\frac{\dot{τ}(t_m^*-t)}{Δσ^*}+1\right)^{2/n} - 1\right]^{1/2} \\ r_A^*(t_p^* < t < t_f) = h\left[\left(\frac{\dot{τ}(t-t_p^*)}{Δσ^*}+1\right)^{2/n} - 1\right]^{1/2} \end{cases} \quad (2)$$

by applying to Eq. (1) the boundary conditions σ($r_Q^*$, *t*) = σ(0, $t_m^*$) = $σ_0^*$ - $Δσ^*$ and σ($r_A^*$, *t*) = σ(0, $t_p^*$) = $σ_0^*$ + $Δσ^*$, respectively. The parameters $t_m^*$ = $t_{mid}$ - $Δσ^*/\dot{τ}$ and $t_p^*$ = $t_{mid}$ + $Δσ^*/\dot{τ}$ represent the times of quiescence disappearance and of activation appearance, respectively, with σ(0, $t_{mid}$) = $σ_0^*$. The background seismicity regime is defined by subtracting the quiescence and activation envelopes $r_A^*(t)$ and $r_Q^*(t)$ from a larger constant envelope $r_{max}$ ≥ max($r^*$) (Fig. 1b). While trivial along $\vec{r}$, concepts of geometric modelling may be required to represent these seismicity solids in three-dimensional Euclidian space [Gallier, 1999] in which the vector $\vec{r}$ is possibly curved [Mignan, 2011]. The non-stationary background seismicity rate μ(*t*) is then defined in the volume of maximum extent $r_{max}$ by

$$μ(t) = \begin{cases} δ_{b0}kr_{max}^d & ,t < t_0 \\ δ_{b0}k(r_{max}^d - r_Q^*(t)^d) + δ_{bm}kr_Q^*(t)^d & ,t_0 \leq t < t_m^* \\ δ_{b0}kr_{max}^d & ,t_m^* \leq t \leq t_p^* \\ δ_{b0}k(r_{max}^d - r_A^*(t)^d) + δ_{bp}kr_A^*(t)^d & ,t_p^* < t < t_f \end{cases} \quad (3)$$

with *k* a geometric parameter and *d* the spatial dimension. For the tectonic case in which $r_{max} \gg h$, the volume is assumed a cylinder with $k = \pi$, $d = 2$ and $\delta$ the density of epicentres in space (Fig. 1c). Finally, the cumulative number of events *N(t)* is defined as

$$N(t) = \int_0^{t_f} \mu(t)dt \qquad (4)$$

which represents a power-law time-to-failure equation of the form

$$N(t) \propto t + t^{\frac{d}{n}+1} \qquad (5)$$

the first term representing the linear background seismicity and the second term the quiescence or activation power-law behaviour observed prior to some large mainshocks (Fig. 1d) [Sammis and Sornette, 2002].

## 3. Application of the N-C PAST static stress model to induced seismicity

In the case of an EGS stimulation, the stress source is the fluid injected at depth with overpressure

$$P(t, r = 0) = K \frac{\Delta V(t, \Delta t)}{V_0} \qquad (6)$$

where *K* is the bulk modulus, $\Delta V$ the volume change per time unit and $V_0$ the infinitesimal volume subjected to pressure effect per time unit at the borehole located at *r* = 0. The injected volume *V(t)* is determined from the flow rate profile *Q(t)*, as

$$V(t) = \int_{t_0}^{t} Q(t)dt \qquad (7)$$

with $t_0$ the starting time of the injection. The change of volume is then defined as

$$\Delta V(t, \Delta t) = \frac{V(t) - V(t - \Delta t)}{\Delta t} \qquad (8)$$

with $\Delta t$ a time increment.

In the EGS case, $r \cong h$ with *h* the borehole depth and induced seismicity defined as hypocentres. The spatiotemporal stress field $\sigma(r,t)$ becomes

$$\sigma(r,t) = \begin{cases} \sigma_0^* & , t < t_0 \\ \frac{r_0^n}{(r+r_0)^n} P(t, r=0) + \sigma_0^* & , t \geq t_0 \end{cases} \qquad (9)$$

with $r$ the distance along the stress field gradient from the borehole, $n = 3$ the spatial diffusion exponent for static stress and $r_0 \to 0$ the infinitesimal radius of volume $V_0 = kr_0^d/t_0$, $t_0 = 1$ being the time unit. Activation represents the case when fluids are injected and quiescence when fluids are ejected (bleed-off). It follows that

$$\begin{cases} r_A^*(t|\Delta V \geq 0) = \left( \frac{r_0^{n-d}}{k} \frac{Kt_0}{\Delta \sigma^*} \Delta V(t) \right)^{1/n} - r_0 \\ r_Q^*(t|\Delta V < 0) = \left( -\frac{r_0^{n-d}}{k} \frac{Kt_0}{\Delta \sigma^*} \Delta V(t) \right)^{1/n} - r_0 \end{cases} \qquad (10)$$

which suggests that the spatiotemporal shape of the induced seismicity envelope depends on the $n$th-root of the flow rate profile $Q(t)$ (with $n = 3$ in the static stress case). This parabolic relationship is similar to the generalized form $r(t) \propto m(t)^{1/d}$ derived from nonlinear poroelasticity in a heterogeneous medium where $m$ is the cumulative mass of injected fluid and $d$ the spatial dimension [Shapiro and Dinske, 2009]. The main difference between the two physical approaches is in the underlying stress field, which is here static and in poroelasticity, dynamic and related to the displacement gradient of the fluid mass [Rudnicki, 1986]. It is trivial to derive Eq. (10) from Eq. (9) while numerous assumptions are necessary to obtain the parabolic form $m(t)^{1/d}$ in nonlinear poroelasticity [Shapiro and Dinske, 2009].

The induced seismicity rate $\mu(t)$ is then defined by Eq. (3) but with $r^*$ from Eq. (10), $k = 4\pi/3$ and $d = 3$, assuming a spherical spatial volume (i.e. isotropic stress field). For the activation phase (i.e. stimulation period), it follows that

$$N(t) \propto \Delta V(t)^{\frac{d}{n}+1} \qquad (11)$$

or

$$N(t) \propto V(t)^{\frac{d}{n}} \qquad (12)$$

The induced seismicity case $d = n = 3$ confirms the linear relationship between cumulative injected volume and cumulative number of induced earthquakes $N(t) \propto V(t)$ previously derived from poroelasticity [e.g., Shapiro and Dinske, 2009]. In contrast with poroelasticity, this second law is a direct consequence of the first. The $d = n$ condition also yields the simplified form of Eq. (10)

$$\begin{cases} r_A^*(t|\Delta V \geq 0) \approx \left(\frac{3}{4\pi} \frac{Kt_0}{\Delta \sigma^*} \Delta V(t)\right)^{1/3} \\ r_Q^*(t|\Delta V < 0) \approx \left(-\frac{3}{4\pi} \frac{Kt_0}{\Delta \sigma^*} \Delta V(t)\right)^{1/3} \end{cases} \quad (13)$$

where the one free parameter is the normalized background stress amplitude range $\widehat{\Delta \sigma^*} = \Delta \sigma^*/(Kt_0)$.

**4. Application to the 2006 Basel EGS induced seismicity sequence**

Figure 2 shows the flow rate $Q(t)$ of injected fluids during the 2006 Basel EGS stimulation experiment [Häring et al., 2008] and the spatiotemporal distribution of relocated induced seismicity [Kraft and Deichmann, 2014] above completeness magnitude $M_c = 0.8$. The injection started at 18:00 on 2 December 2006 ($t_0$) and stopped at 11:33 on 8 December 2006 ($t_1$) after which the well was bled-off ($\Delta V < 0$) (Fig. 2a). The N-C PAST thus predicts an activation envelope $r_A^*$ for $t_0 \leq t < t_1$ and a quiescence envelope $r_Q^*$ for $t \geq t_1$ (Eq. 13). The activation and quiescence envelopes are fitted to the Basel data using $\widehat{\Delta \sigma^*} \in [10^{-3}, 10^{-1}]$ day$^{-1}$ (light curves) and $\Delta t = 1/4$ day. The results are shown in Figure 2b. The value $\widehat{\Delta \sigma^*} = 0.007$ day$^{-1}$ (dark curves) provides the best fit to the data, defined from the best score $S = (w_A + w_Q)/2$ with $w_A$ and $w_Q$ the ratio of events of distance $r \leq r_A^*$ and $r \geq r_Q^*$ in the injection and bleeding-off phases, respectively. Figure 2c shows $S$ as a function of $\widehat{\Delta \sigma^*}$ for $\Delta t = \{1/12, 1/8, 1/4\}$ day, which indicates that the results remain stable for lower time increments.

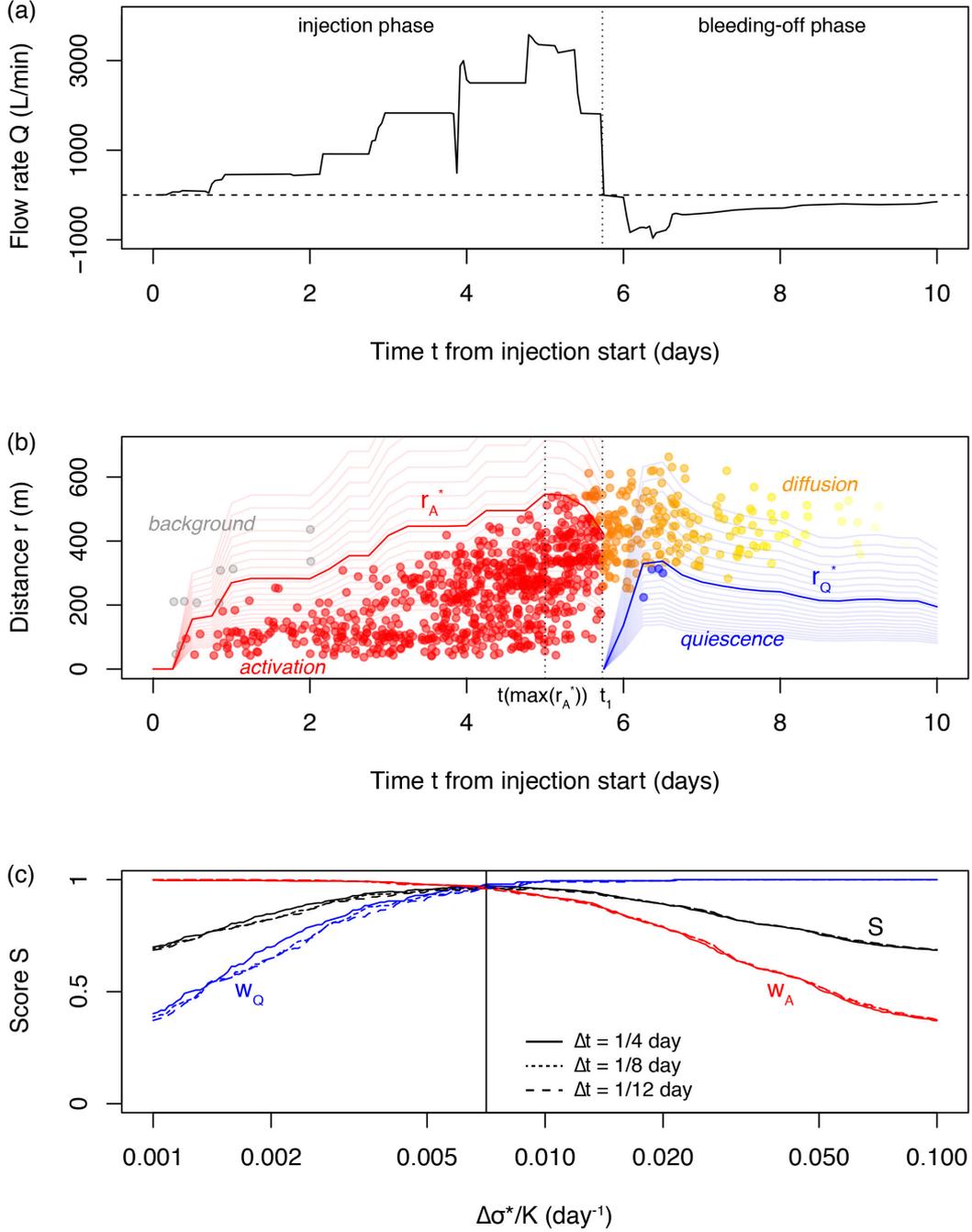

**Figure 2:** 2006 Basel EGS stimulation experiment data with activation and quiescence envelope fits: (a) Flow rate $Q(t)$ [Häring et al., 2008]; (b) Spatiotemporal distribution of relocated induced seismicity [Kraft and Deichmann, 2014] with $r$ the distance from the borehole. The activation and quiescence envelopes $r_A^*(t)$ and $r_Q^*(t)$ are defined from Eq. (13) with parameters $\widehat{\Delta\sigma^*} = 0.007$ day$^{-1}$ (dark curves) and $\Delta t = 1/4$ day. The light curves represent the range $\widehat{\Delta\sigma^*} \in [10^{-3}, 10^{-1}]$ day$^{-1}$ in 0.1 increments

in the $\log_{10}$ scale. Points represent the induced earthquakes, which colour indicates how they are declared; (c) Score $S = (w_A+w_Q)/2$ with $w_A$ and $w_Q$ the ratio of events of distance $r \leq r_A^*$ and $r \geq r_Q^*$ in the injection and bleeding-off phases, respectively. The vertical line represents $\widehat{\Delta\sigma}^* = 0.007$ day$^{-1}$.

I evaluate $\delta_{b0} = 10^{-10}$ event/m$^3$/day by counting all earthquakes declared in the national Swiss catalogue (ECOS-09[1]) and located within 10 km of the borehole of coordinates (7.594°E; 47.586°N) and depth 4.36 km. It means that ~1 tectonic earthquake is expected in average in the space-time window considered. Due to the low tectonic activity in the area, I approximate $\delta_{b0} = \delta_{bm} = 0$ event/m$^3$/day (i.e., total quiescence). The theory shows a good agreement with the observations with 97% of the seismicity below $r_A^*$ during the injection phase (red points in Fig. 2b) and 98% of the seismicity above $r_Q^*$ during the bleeding-off phase (orange to yellow points).

The density of events above $r_Q^*$ is however not $\delta_{b0}$ but

$$\delta_b(t \geq t_1) = \delta_{bp} \exp\left(-\frac{t-t_1}{\tau}\right) \qquad (14)$$

which represents the temporal diffusion of induced seismicity with $\tau$ the average time constant. Eq. (14) represents a relaxation process from the overloading state to the background state. The results here suggest that only the events declared as background (grey points) and quiescence events (blue points) are outliers. The observed variations in $r$ below $r_A^*$ and above $r_Q^*$ are not explained by the model, which only predicts the behaviour of the activation and quiescence fronts. The second-order variations may be due to anisotropic effects and for $t > t(\max(r_A^*))$ to additional spatial diffusion effects.

---

[1] http://hitseddb.ethz.ch:8080/ecos09/

Figure 3 shows the 6-hour rate of induced seismicity μ(*t*) and the cumulative number of induced events *N(t)*, observed and predicted. With $\delta_{b0} = \delta_{bm} = 0$ and taking into account induced seismicity temporal diffusion, the rate of induced seismicity becomes

$$\mu(t) = \max\left(\frac{4\pi}{3}\delta_{bp}.\Delta t.r^*(t)^3, \frac{4\pi}{3}\delta_{bp}.\Delta t.r^*(t-S_t)^3 \exp\left(-\frac{t-S_t}{\tau}\right)\right) \quad (15)$$

where $\delta_{bp}$ = 4.68 10$^{-7}$ event/m$^3$/day (production parameter) and τ = 1.18 day (diffusion parameter) are obtained by maximum-likelihood estimation (MLE), set $S_t$ = {Δ*t*, …, *i*Δ*t*, …} and

$$r^*(t) = \begin{cases} 0 & , t < t_0 \\ r_A^*(t) & , t_0 \leq t < t_1 \\ 0 & , t \geq t_1 \end{cases} \quad (16)$$

Eq. (15) infers that induced seismicity is fully explained by overloading, in agreement with the observation of no causal relationships between events in the Basel sequence [Langenbruch et al., 2011]. The predicted rate (Eq. 15) and predicted cumulative number of events (Eq. 4) fit the data well, as shown in Figures 3a and 3b, respectively. The role of temporal diffusion is observed after $t_1$-Δ*t* and is the only contributor to induced seismicity after $t_1$. Of three functional forms tested to describe diffusion (exponential, stretched exponential and power law), the exponential (Eq. 14) was verified to be the best model for the Basel case (following the formalism and tests proposed by Clauset et al. [2009]).

## 5. Conclusions

I have demonstrated that the two principal induced seismicity descriptive laws can be explained from simple geometric operations in a static stress field without requiring any concept derived from poroelasticity. The two descriptive laws had been previously obtained by considering the differential equations of poroelasticity [Biot, 1941; Rudnicki, 1986] under different assumptions [Shapiro and Dinske, 2009], which indicates that the static stress model defined from algebraic expressions requires a lower description length [Kolmogorov, 1965]. This is crudely inferred here from the difference between the lengths of the present demonstration and of published poroelasticity demonstrations [e.g., Shapiro and Dinske, 2009].

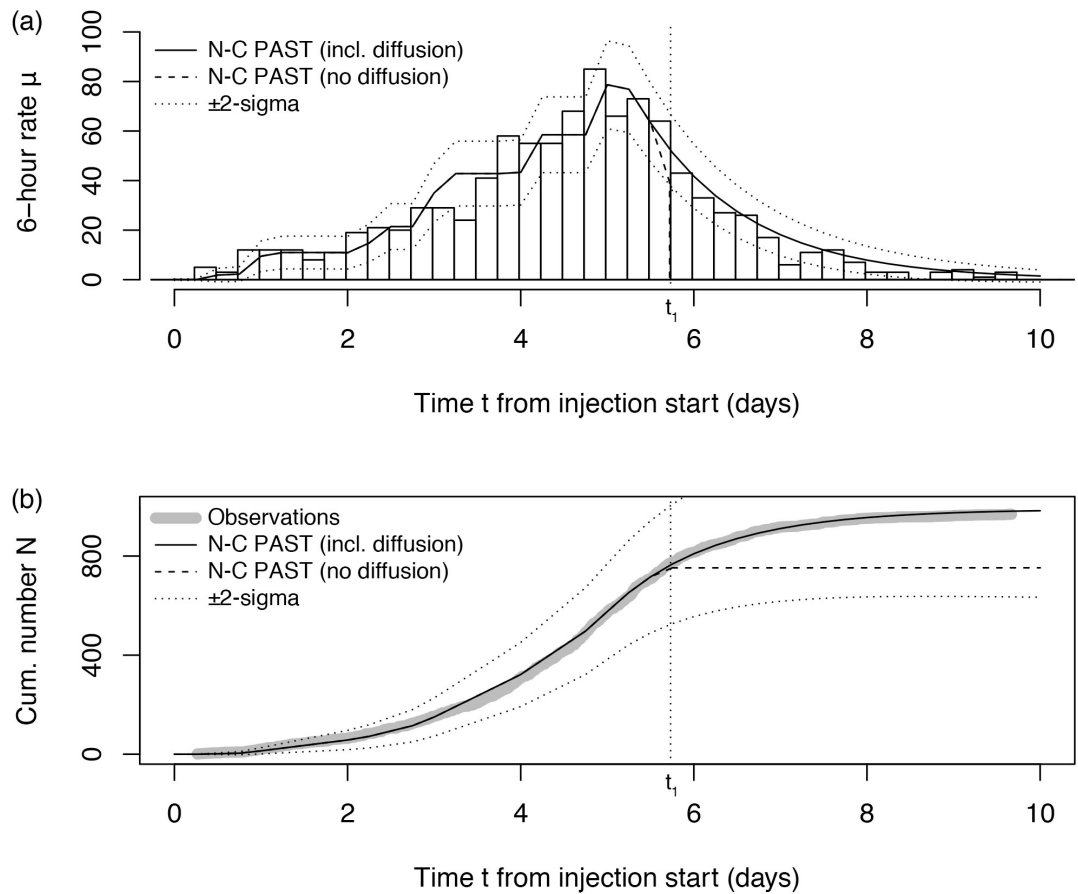

**Figure 3:** Induced seismicity production time series, observed and predicted: (a) Histogram of the observed 6-hour induced seismicity rate µ(*t*) with fit based on Eq. (15) with MLE parameters $\delta_{bp}$ = 4.68 $10^{-7}$ event/m$^3$/day (production parameter) and τ = 1.18 day (diffusion parameter); (b) Cumulative number of induced earthquakes *N*(*t*) with fit based on Eq. (4) with µ(*t*) of Eq. (15).

I also showed that the controlling parameter is the normalized background stress amplitude range $\widehat{\Delta\sigma^*}$, which questions the usefulness of permeability and diffusivity parameters in induced seismicity analyses and might explain why these parameters remain elusive [Miller, 2015]. In that view, permeability could depend on the "external loading configuration" instead of on the material itself, as recently proposed in the case of the static friction coefficient [Ben-David and Fineberg, 2013]. Testing of the model on other induced seismicity sequences will determine if $\widehat{\Delta\sigma^*}$ is itself universal, region-specific or related to the static stress memory of the crust, hence if $\widehat{\Delta\sigma^*}$ depends or not on the tectonic loading configuration at EGS natural laboratory sites. Similar questions apply to the earthquake production parameter $\delta_{bp}$ and if the two parameters are independent or correlated.

The main assumption of the N-C PAST is to consider three unique seismicity regimes (quiescence, background and activation) defined by the event productions $\delta_{bm} < \delta_{b0} < \delta_{bp}$. There are two possible physical alternatives to justify this choice: (1) it represents the fundamental behaviour of the Earth crust, which would hence act as a capacitor, with strain energy storage and $\delta_{bp}$ analogues to electrical energy storage and capacitance, respectively; (2) the proposed step function is a simplification of the true stress-production profile, which remains unknown and is so far best characterized by three regimes [e.g., King, 2007]. Both alternatives allow defining spatiotemporal solids over which geometrical operations yield algebraic expressions of the induced seismicity behaviour.